\documentclass[10pt,journal,compsoc]{IEEEtran}
\usepackage{multirow}
\usepackage{graphicx}
\usepackage{caption}
\usepackage{url}
\usepackage{float}

\ifCLASSOPTIONcompsoc
  \usepackage[nocompress]{cite}
\else
  \usepackage{cite}
\fi
\ifCLASSINFOpdf
\else
\fi
\begin{document}
%
\title{How COVID-19 Has Changed Crowdfunding: Evidence From GoFundMe}

\author{Junda Wang, Xupin Zhang*,
        and Jiebo Luo*,~\IEEEmembership{Fellow,~IEEE} 

\IEEEcompsocitemizethanks{\IEEEcompsocthanksitem  Junda Wang is with the Department of Computer Science, University of Rochester, Rochester, NY 14627.\protect\\
E-mail: jwang212@ur.rochester.edu

\IEEEcompsocthanksitem  Xupin Zhang is with the Department of Human Development, University of Rochester, Rochester, NY 14627.\protect\\
E-mail: xzhang72@u.rochester.edu

\IEEEcompsocthanksitem Jiebo Luo is with the Department of Computer Science, University of Rochester, Rochester, NY 14627.\protect\\
E-mail: jluo@cs.rochester.edu
}
}


%


\maketitle

\begin{abstract}
While the long-term effects of COVID-19 are yet to be determined, its immediate impact on crowdfunding is nonetheless significant. This study takes a computational approach to more deeply comprehend this change. Using a unique data set of all the campaigns published over the past two years on GoFundMe, we explore the factors that have led to the successful funding of a crowdfunding project. In particular, we study a corpus of crowdfunded projects, analyzing cover  images and other variables commonly present on crowdfunding sites. Furthermore, we construct a classifier and a regression model to assess the significance of features based on XGBoost. In addition, we employ counterfactual analysis to investigate the causality between features and the success of crowdfunding. More importantly, sentiment analysis and the paired sample t-test are performed to examine the differences in crowdfunding campaigns before and after the COVID-19 outbreak that started in March 2020. First, we note that there is significant racial disparity in crowdfunding success. Second, we find that sad emotion expressed through the campaign's description became significant after the COVID-19 outbreak. Considering all these factors, our findings shed light on the impact of COVID-19 on crowdfunding campaigns.
\end{abstract}


\begin{IEEEkeywords}
crowdfunding, GoFundMe, counterfactual, topic model, XGBoost
\end{IEEEkeywords}

%
\IEEEpeerreviewmaketitle

\section{Introduction}

In recent years, with the development of the Internet, there are more ways to raise money online. GoFundMe, an American for-profit crowdfunding platform that encourages people to create online crowdfunding projects about their life events such as illnesses and accidents, is a good example. Even if crowdfunding becomes more and more convenient, the success rate of crowdfunding is still not high. To date, there is not much information regarding a recipe for a successful campaign, and the factors leading to the success of crowdfunding have become a critical research goal\cite{kaartemo2017elements}. 

In the current study, we analyze GofundMe to extract the most critical factors that contribute to crowdfunding success. We collect data of 36,370 campaigns on GoFundMe and split them into two parts. One part of the crowdfunding data is for before the COVID-19 outbreak (2019), and the other is for after the COVID-19 outbreak (2020). In this regard, we also analyze whether the influencing factors have changed under the influence of the pandemic. Some researchers believed the emotional elements conveyed through text and facial expressions are likely to attract donors\cite{rhue2018emotional}. We extract the face from the picture and judge the emotion using the Baidu Application Programming Interface (API)\footnote{https://cloud.baidu.com/doc/FACE/s/Uk37c1m9b\label{web}}. In addition, we extract and infer individual-level features such as gender, race, age, beauty, target, location, followers, shares, distinct donors, family status, and facial attractiveness, as well as duration of crowdfunding. Text is one of the most basic aspects of information transfer, and some researchers believed that textual elements including descriptions, reviews, and emotion have a considerable impact on the success of crowdfunding\cite{koch2019recipe}.  Therefore, we incorporate text emotion into models through a text scoring model that produces the scores as a feature. According to the characteristics mentioned above, a comprehensive and in-depth analysis can be performed. If a campaign page visitor’s sympathy with specific project topics is given, a more longer visit duration would increase the probability that the page visitor donates and thus the success of the crowdfunding \cite{koch2019recipe}. Hence, the aesthetic and technical scores of the cover image are also considered as factors affecting the success of crowdfunding\cite{zhang2020contributes}. Based on the above data, we propose three hypotheses:

\begin{itemize}
    \item \textbf{Hypothesis 1:} The basic features of crowdfunding, fundraiser and descriptions have a significant impact on fund-raising success.
    \item \textbf{Hypothesis 2:} There are significant differences between before and after the COVID-19 outbreak.
    \item \textbf{Hypothesis 3:} There are certain social disparities related to different races and the compound factors, which consist of the impact of COVID-19 and other social factors.
\end{itemize}

The goal of this study focuses on analyzing the main factors that affect  crowdfunding campaigns and the impact of the COVID-19 on these factors. Notably, we define a success group where the crowdfunding projects have raised  more than the target amounts, and a failure group otherwise. Moreover, we build predictions models for both classification of success/failure and regression of the raised amount. For regression or classification problems, we use XGBoost to solve these two problems and provide the list of essential factors. Finally, we perform a counterfactual experiment to analyze the influences of different factors and the significant impact of the COVID-19 on these factors. To our best knowledge, this is the first study examining the impact of COVID-19 on crowdfunding campaigns.

\section{Related Work}

In the past few years, more platforms have been created to provide online crowdfunding, such as Kickstarter, Indiegogo, and GoFundMe. Particularly, crowdfunding platforms aim to be intermediaries between sponsors and fundraisers. They can also promote the ideas of the fundraiser and call for support from potential investors\cite{belleflamme2014crowdfunding}. In addition to the factors mentioned above, the funding goal\cite{frydrych2014exploring} and project duration\cite{barbi2017crowdfunding} have a considerable impact on the success of crowdfunding.

Increasingly more studies have investigated crowdfunding success using data-mining methods. Specifically, they  predict crowdfunding success and analyze the factors that are significant to crowdfunding. Mollick argued that it is difficult to measure success and performed a manual analysis to distinguish between success and failure\cite{mollick2014dynamics}. Yuan  identified the importance of topics and sentiment further affecting the success of crowdfunding\cite{yuan2016determinants}. Notably, Mitra and Gilbert predicted the success of crowdfunding through analyzing text using LIWC\cite{mitra2014language}. Lauren examined the relationship between the emotion of the cover image and the crowdfunding success\cite{rhue2018emotional} and found that crowdfunding platforms rely on emotions to attract sponsors.

\section{Data and Features}

\subsection{Data sets}

We  focus on GofundMe to analyze the crucial factors contributing to the success. Specifically, we crawl all the available 36,370 crowdfunding campaigns on GofundMe  and divide them into two parts. One part was collected before August 2019, while the other was collected after August 2020. The purpose of this division is to analyze whether COVID-19 has had an impact on people’s attitude towards crowdfunding or whether some influential factors or less influential factors have changed. The features of the dataset are summarized below and shown in Table ~\ref{feature}. 

\subsection{Basic Features}
The features that can be directly extracted from the website are as follows: launch date, cover image, description, category, current amount, goal amount, \# of followers, \# of shares, and \# of donors. We refer to these features as the basic features.

\subsection{Inferred Features}

\begin{itemize}
\item \textbf{Quality Scores:} We use the pre-trained model NIMA to obtain the aesthetic and technical scores of each cover image\cite{talebi2018nima}.

\item \textbf{Text Features:} First, we merge the titles and descriptions of the campaigns and use them as our text data. Next, we employ Vader~\cite{hutto2014vader} to evaluate individual text data and obtain three predicted emotion scores including positive, negative, and neutral scores, respectively. Finally, we examine the text using LIWC to obtain more detailed text sentiment scores: sad scores, anger scores, and anxiety scores. However, it is evident that both descriptions of the campaigns are essential in impacting crowdfunding success. These potential effects are not easy to measure directly, therefore we train an XLNet model to predict the success of the crowdfunding campaign and learn its potential effects\cite{yang2019xlnet}.

\item \textbf{Image Features:} In terms of image features, we compare the DeepFace API~\cite{serengil2020lightface} with the Baidu API, and also consider the previous research comparing the Baidu API and  other competing APIs~\cite{yang2020benchmarking}. In the end, we choose the Baidu API, which is a platform providing reliable face recognition services. The Baidu API returns facial attractiveness, age, race, emotion, and gender of each face in the cover image (also referred to as profile image). For simplicity, we calculate the mean attractiveness of faces when an image contains more than one person. For the age feature, we not only calculate the mean age but also return two characteristics: the number of children and the number of older adults. People under 15 years old are defined as children, while those over 60 years are regarded as older adults. We regard the number of people of different races as the characteristics variable. In the Baidu API, there are four different races: Black, White, Asian, and Others. In terms of gender, we obtain the number of males and females. In the end, we extract the emotion of each face (happy, sad, grimace, neutral, or angry).

\end{itemize}

\begin{table*}[!htbp] \centering 
  \caption{Sources of features.} 
  \label{feature} 
\begin{tabular}{@{\extracolsep{5pt}}lccccccc} 
\\[-1.8ex]\hline 
\hline \\[-1.8ex] 
Statistic &  \multicolumn{1}{c}{Mean} & \multicolumn{1}{c}{St. Dev.} & \multicolumn{1}{c}{Source}\\ 
\hline \\[-1.8ex] 
\textbf{basic features}\\ 
Goal  & 176,738.900 & 10,635,100.000 & Web crawler\\ 
Shares  & 844.223 & 2,439.918 & Web crawler\\ 
Donors  & 235.444 & 1,190.212 & Web crawler\\ 
Followers  & 234.578 & 1,153.083 & Web crawler\\ 
Category & & & Web crawler\\
days  & 322.139 & 288.000 & Manually coded\\ 
\hline 
\textbf{text features}\\ 
text\_positive  & 0.175 & 0.071 &Vader\\ 
text\_neutral  & 0.779 & 0.076 &Vader\\ 
text\_negative  & 0.046 & 0.042 &Vader\\ 
anx  & 0.178 & 0.353 &LIWC2015\\ 
anger  & 0.227 & 0.509 &LIWC2015\\ 
text\_sad  & 0.487 & 0.817 &LIWC2015\\ 
text\_scores  & $-$0.501 & 0.468 &XLnet\\ 
\hline 
\textbf{image features}\\
age  & 27.235 & 10.952 &Average of Baidu API\\ 
have\_kid  & 0.534 & 1.159 &Manually coded(by age feature)\\ 
old  & 0.028 & 0.179 &Manually coded(by age feature)\\ 
facial attractiveness  & 35.963 & 12.408 &the sum of Baidu API's result\\ 
Black  & 0.382 & 1.346 &the sum of Baidu API's result\\ 
Asian  & 0.709 & 1.600 &the sum of Baidu API's result\\ 
White  & 2.062 & 2.725 &the sum of Baidu API's result\\ 
other  & 0.025 & 0.205 &the sum of Baidu API's result\\ 
happy  & 2.003 & 2.883 &the sum of Baidu API's result\\ 
sad  & 0.192 & 0.516 &the sum of Baidu API's result\\ 
grimace  & 0.012 & 0.113 &the sum of Baidu API's result\\ 
neutral  & 0.673 & 1.670 &the sum of Baidu API's result\\ 
angry  & 0.050 & 0.262 &the sum of Baidu API's result\\ 
male  & 1.819 & 2.703 &the sum of Baidu API's result\\ 
female  & 1.359 & 2.129 &the sum of Baidu API's result\\ 
\hline \\[-1.8ex] 
\end{tabular} 
\end{table*}

\section{Methodology and Findings}

\subsection{Campaign Category}

\begin{figure*}
	\includegraphics[width=1\textwidth]{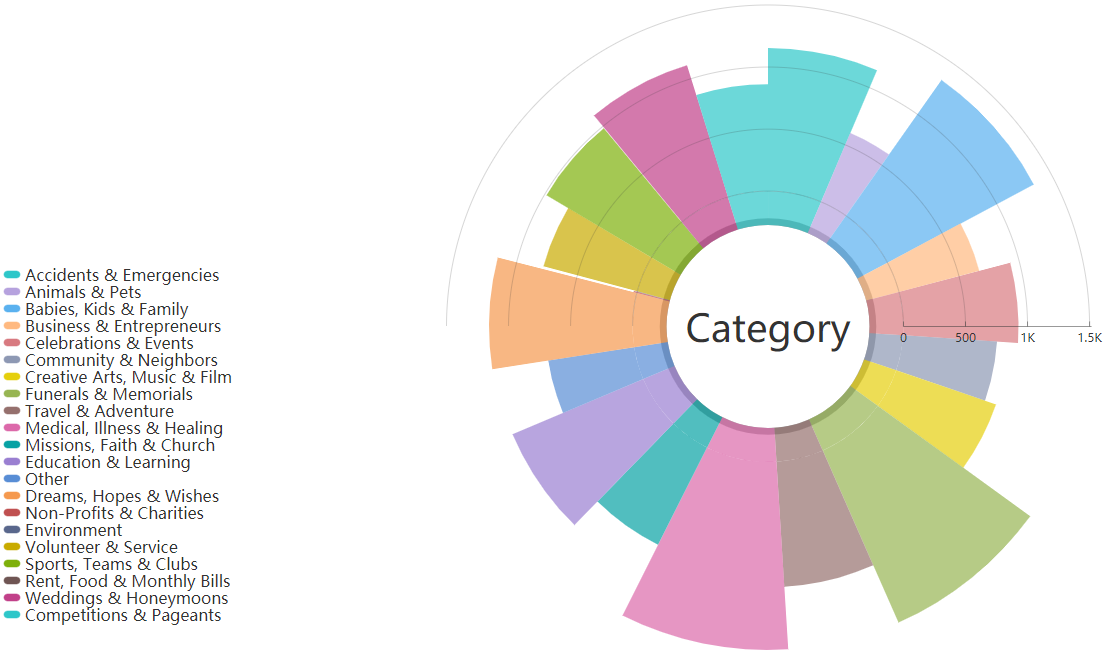}
	\centering
	\caption{Category distribution.}
	\label{dcategory}
\end{figure*}

There are 21 different categories of campaigns and their distribution is shown in  Figure~\ref{dcategory}. Some of these categories are related to the success rate of crowdfunding, while others are not. Specifically, we analyze the categories that have the most significant impact on crowdfunding success using the t-tests. We find that the impact of some categories on crowdfunding success has changed over the past two years. Since it makes no sense to compare p values directly, we compare whether their significance levels has changed. For example, the significance levels of the \emph{Travel \& Adventure}, \emph{Non-Profits \& Charities}, \emph{Medical, Illness \& Healing}, \emph{Celebrations \& Events}, \emph{Creative Arts, Music \& Film}, \emph{Accidents \& Emergencies} and \emph{Dreams, Hops \& Wishes} categories have changed. In summary, the success ratios of these categories have increased, except for the \emph{Travel \& Adventure} category. After the COVID-19 outbreak, it is evident that people are even more reluctant to donate to crowdfunding in the \emph{Travel \& Adventure} category. In particular, the \emph{Medical, Illness \& Healing} and \emph{Non-Profits \& Charities} categories were not significant before the COVID-19 outbreak but have become very significant after the COVID-19 outbreak. For the \emph{Dreams, Hops \& Wishes} category, people who were more willing to donate are now reluctant. For other declining categories, it is evident that in most of the categories that people were not willing to donate money to before, people have become even more reluctant to donate to later, such as \emph{Sports, Teams \& Clubs}, \emph{Missions, Faith \& Church}, \emph{Weddings \& Honeymoons} and \emph{Competitions \& Pageants}. Finally and more significantly,  the success rate of every category has declined after the COVID-19 outbreak.

\begin{table} \centering 
  \caption{Paired sample statistics.} 
  \label{category} 
\begin{tabular}{@{\extracolsep{5pt}}lccc} 
\\[-1.8ex]\hline 
\hline \\[-1.8ex] 
 & \multicolumn{2}{c}{\textit{Dependent variable:}} \\ 
\cline{2-3} 
\\[-1.8ex] & \multicolumn{2}{c}{Year} \\ 
\\[-1.8ex] & 2020 & 2019 &Ratio\\ 
\hline \\[-1.8ex] 
 Travel \& Adventure                               & -0.082$^{***}$  & -0.02$^{*}$ & -0.015\\
 Environment                                      & 0.014      &     &    \\
 Babies, Kids \& Family       & 0.035$^{***}$  & 0.044$^{***}$ &0.014 \\
 Sports, Teams \& Clubs       & -0.040$^{***}$ & -0.048$^{***}$ &-0.055\\
 Competitions \& Pageants     & -0.092$^{***}$  & -0.090$^{***}$ &  -0.136\\
 Non-Profits \& Charities     & 0.019$^{**}$    & 0.007 & 0.003\\
 Medical, Illness \& Healing  & 0.046$^{***}$  & 0.016 & 0.004\\
 Volunteer \& Service         & 0.010      & 0.011 & -0.012\\
 Business \& Entrepreneurs    & -0.043$^{***}$  & -0.080$^{***}$ & 0.005\\
 Weddings \& Honeymoons       & -0.099$^{***}$ & -0.051$^{***}$ & 0.007  \\
 Funerals \& Memorials        & 0.1254$^{***}$ & 0.119$^{***}$ & 0.011 \\
 Missions, Faith \& Church    & -0.043$^{***}$  & -0.041$^{***}$ & -0.026  \\
 Education \& Learning        & 0.001      & 0.011 & -0.003      \\
 Animals \& Pets              & -0.002      & -0.007 & 0.008\\
 Celebrations \& Events       & -0.031$^{**}$  & -0.031$^{*}$ & 0.002 \\ 
 Creative Arts, Music \& Film & -0.035$^{***}$  & -0.003 & 0.013\\
 Accidents \& Emergencies     & 0.052$^{***}$  & 0.031$^{***}$ & 0.016    \\
 Dreams, Hopes \& Wishes      & -0.001      & 0.019$^{*}$ & 0.016     \\ 
 Rent, Food \& Monthly Bills  & 0.013      &  &          \\
 Other                       & 0.029$^{*}$     & 0.014 & 0.190      \\
 Community \& Neighbors       & 0.023    & 0.033$^{**}$ & -0.003    \\
 ALL       &     &  & -0.003    \\
\hline 
\hline \\[-1.8ex] 
\textit{Note:}  & \multicolumn{2}{r}{$^{*}$p$<$0.1; $^{**}$p$<$0.05; $^{***}$p$<$0.01} \\ 
\end{tabular} 
\end{table}

\subsection{Prediction}

To analyze the contribution of influential factors to success, we employ the XGBoost method and divide features into three categories: basic features, text features, and image features. We feed (1) basic features, (2) basic features plus text features, and (3) all the features into an XGBoost model to obtain the accuracy, F1 score, precision, and recall, respectively. We also train an XGBoost model to construct a regression model with the ratio/percentage of success as the output. It is evident that the inferred features significantly improve the model performance  (Table~\ref{resultx}). 

To extract statistically significant features, we construct a logistic regression model for the classification problem. The result is shown in Table~\ref{lg} and validates {\bf Hypothesis 1}. To further analyze the impact of each feature on the success of crowdfunding, we conduct several counterfactual analysis experiments. Since we have too many features, we analyze the statistically significant features in a separate analysis.

\begin{table*}\centering
    \caption{XGBoost performance metrics.}
    \label{resultx}
	\begin{tabular}{cccccc}
		\hline
		& \multicolumn{4}{c}{Classification}         & Regression  \\ \hline \hline
		& Accuracy & F1       & Precision & Recall   & R-square \\ \hline
		Base      & 0.823945 & 0.701222 & 0.756329  & 0.653601 & 0.28585  \\ \hline
		Base+Text & 0.835357 & 0.72435  & 0.769849  & 0.683929 & 0.49147  \\ \hline
		All       & 0.852302 & 0.757083 & 0.8009335  & 0.711169 & 0.54452  \\ \hline
	\end{tabular}
\end{table*}

\begin{table*}[!htbp] \centering 
  \caption{Logistic regression model results for the relationship between campaigns' features and success} 
  \label{lg} 
\begin{tabular}{@{\extracolsep{5pt}}lcccc} 
\\[-1.8ex]\hline 
\hline \\[-1.8ex] 
 & \multicolumn{4}{c}{\textit{Dependent variable: success}} \\ 
\cline{2-5} 
\\[-1.8ex] & \multicolumn{4}{c}{models} \\ 
\\[-1.8ex] & basic model & basic+text model & text+image model & aggregated model\\ 
\hline \\[-1.8ex] 
 Goal & $-$0.00005$^{***}$ (0.00000) & $-$0.00005$^{***}$ (0.00000) &  & $-$0.00005$^{***}$ (0.00000) \\ 
  Shares & 0.00004$^{***}$ (0.00001) & 0.00002$^{*}$ (0.00001) &  & 0.00002 (0.00001) \\ 
  Donors & 0.004$^{***}$ (0.0004) & 0.003$^{***}$ (0.0004) &  & 0.003$^{***}$ (0.0004) \\ 
  Followers & $-$0.0003 (0.0003) & $-$0.0005 (0.0004) &  & $-$0.001 (0.0004) \\ 
  days & $-$0.00002 (0.0001) & $-$0.001$^{***}$ (0.0001) &  & $-$0.001$^{***}$ (0.0001) \\ 
  text\_positive &  & 44.048 (39.045) & 39.543 (35.766) & 46.768 (39.212) \\ 
  text\_neutral &  & 43.616 (39.044) & 39.151 (35.766) & 46.565 (39.211) \\ 
  text\_negative &  & 44.307 (39.045) & 38.454 (35.766) & 47.058 (39.212) \\ 
  anx &  & $-$0.105$^{**}$ (0.051) & $-$0.060 (0.048) & $-$0.097$^{*}$ (0.052) \\ 
  anger &  & 0.027 (0.036) & 0.050 (0.034) & 0.031 (0.037) \\ 
  text\_sad &  & $-$0.001 (0.023) & 0.023 (0.021) & $-$0.006$^{*}$ (0.023) \\ 
  text\_scores &  & 2.077$^{***}$ (0.049) & 1.929$^{***}$ (0.045) & 2.055$^{***}$ (0.049) \\ 
  age &  &  & $-$0.005$^{***}$ (0.002) & $-$0.0001 (0.002) \\ 
  have\_kid &  &  & $-$0.030 (0.020) & 0.038$^{*}$ (0.020) \\ 
  old &  &  & 0.040 (0.100) & $-$0.049 (0.111) \\ 
  facial attractiveness &  &  & $-$0.002 (0.001) & $-$0.00003 (0.002) \\ 
  Black &  &  & $-$0.052 (0.034) & $-$0.078$^{**}$ (0.036) \\ 
  Asian &  &  & $-$0.064$^{**}$ (0.033) & $-$0.065$^{**}$ (0.035) \\ 
  White &  &  & $-$0.043 (0.031) & $-$0.018 (0.034) \\ 
  other &  &  & $-$0.257$^{**}$ (0.102) & $-$0.152 (0.112) \\ 
  happy &  &  & 0.037 (0.031) & 0.058$^{*}$ (0.033) \\ 
  sad &  &  & 0.050 (0.048) & 0.046 (0.052) \\ 
  grimace &  &  & 0.182 (0.151) & 0.153 (0.164) \\ 
  neutral &  &  & 0.036 (0.035) & 0.020 (0.037) \\ 
  angry &  &  & $-$0.056 (0.081) & $-$0.095 (0.087) \\ 
  female &  &  & 0.005 (0.014) & $-$0.021 (0.015) \\ 
  male &  &  &  &  \\ 
  aesthetic scores &  &  & 0.003 (0.032) & $-$0.027 (0.035) \\ 
  technical scores &  &  & $-$0.009 (0.032) & 0.143$^{***}$ (0.036) \\ 
  Constant & $-$0.417$^{***}$ (0.028) & $-$42.840 (39.043) & $-$38.840 (35.766) & $-$46.402 (39.211) \\ 
 \hline \\[-1.8ex] 
Observations & 19,185 & 19,185 & 19,075 & 19,075 \\ 
Log Likelihood & $-$10,364.040 & $-$9,166.808 & $-$10,626.380 & $-$9,080.760 \\ 
Akaike Inf. Crit. & 20,740.080 & 18,359.620 & 21,300.760 & 18,219.520 \\ 
\hline 
\hline \\[-1.8ex] 
\textit{Note:}  & \multicolumn{4}{r}{$^{*}$p$<$0.1; $^{**}$p$<$0.05; $^{***}$p$<$0.01} \\ 
\end{tabular} 
\end{table*}

\section{Counterfactual analysis}

We compare the impact of the above mentioned features on crowdfunding success before and after the outbreak of COVID-19 and further analyze whether the COVID-19 pandemic has changed people's concerns and priorities. To have a better understanding of the causality between the aforementioned features and crowdfunding success rate, we employ counterfactual analysis, which has been widely used in comparative inquiry\cite{chang2021mobility}. We test the remaining two of the three hypotheses formulated in the beginning based on the results of the counterfactual analysis experiments.

\noindent\textbf{Removing the category factors to reduce the impact} \quad In counterfactual analysis, for each category, we turn the factor indicator of each category into 0 in each experiment as shown in Table~\ref{ccategory}. We find that \emph{Creative Arts, Music \& Film} and \emph{Dreams Hopes \& Wishes} have positive causal impacts on the success of crowdfunding before COVID-19 while having a negative causal effect after the COVID-19 outbreak. The positive effect of \emph{Medical, Illness \& Healing}, \emph{Accidents \& Emergencies} and \emph{Non-Profits \& Charities} on the success of crowdfunding has dramatically increased after the COVID-19 outbreak. The significance of these three categories in Table~\ref{category} has changed, indicating that the impact of COVID-19 on these three categories is profound. This shows that the COVID-19 pandemic has changed individuals' preferences and individuals focus more on medical or charity topics than dreams, arts and charity topics. These results validate {\bf Hypothesis 2}. It is worth mentioning that the ratio in the table changes little because of the multiple kinds of categories. Eliminating the impact of one of these categories cannot significantly change the overall success, but it does not mean that these categories are not significant.

\noindent\textbf{Improving the effect of features} First, we divide the remaining features into the basic features, text features, and image features after removing the category features. Combined with the counterfactual experiment, we focus on the analysis of those statistically significant features. Table~\ref{fcategory} shows the result of counterfactual analyses of statistically significant features using logistic regression. When increasing these features with their mean values, the percentage of success increases significantly.

\begin{itemize}
\item \textbf{Goal Amount} \quad From the results of the counterfactual experiment, the feature with the most significant causal relationship with success is the goal amount. The displayed funding goal functions is a signal of difficulty/complexity of the project. In general, the higher the crowdfunding goal is, the lower the success rate of the project is\cite{barbi2017crowdfunding}. Our counterfactual experiment results are also consistent with the previous results. 

\item \textbf{Duration} \quad According to the experimental results, we find that the duration of crowdfunding campaigns has a positive impact on the success. The longer a crowdfunding project lasts, the more likely people have the opportunity to find and share it ($p<0.001$). The campaigns will receive more donations with an increased number of shares ($p<0.001$). 

\item \textbf{Donors} \quad The relationship between the number of donors and success is significant ($p<0.001$). The number of donors has a positive impact on crowdfunding success as shown by the counterfactual experiment results.

\item \textbf{Emotion of text} \quad Text scores obtained by XLNet have a significant positive impact on crowdfunding success since the outbreak of COVID-19. It implies that there is a statistically significant relationship between text and success. Further analysis shows that there is a significant relationship between emotion and success rate. For example, anxiety ($p<0.1$) and sadness ($p<0.1$) of text have a negative impact on success. This result is consistent with our counterfactual experiment.

\item \textbf{Technical Score} \quad Technical scores obtained by Pretrained model NIMA have a significant positive impact on crowdfunding success since the outbreak of COVID-19 ($p<0.01$). It implies that there is a statistically significant relationship between the quality of the images and success.

\item \textbf{Race} \quad For image features, we find that race has different levels of influence on the success of crowdfunding since the outbreak of COVID-19. Among all races, Black has the most significant impact on the success ($p<0.05$). It seems that while historically black people are less likely to be funded than other races, there is less significant effect before COVID-19 ($p<0.1$). In other words, black people are even more less likely to be funded during the COVID-19 pandemic. In the whole data set, Asian also has a statistically significant impact on the success of crowdfunding ($p<0.05$) after the COVID-19 outbreak, while before the COVID-19 outbreak, there was no statistically significant impact. These results support {\bf Hypothesis 3}. We suspect that the former might be related to the root cause of the \#BlackLivesMatter movement and the latter might be related to the root cause of the \#StopAsianHate movement. These are two interesting directions to investigate in depth in future work. 

\item \textbf{Emotion of cover image} \quad We examine the relationship  between the emotion of the cover image  and success. We find that there is a significant difference between emotion and success. According to the counterfactual experiments, sadness has a negative effect on success, while happiness has a positive effect on success ($p<0.1$). People are more willing to donate to those whose cover images carry positive valence than those whose images carry negative valence. This is an interesting and somewhat surprising discovery as donors seem to like those who are upbeat and optimistic about life.  

\item \textbf{Children} \quad From the results of the logistic regression model, we find that the impact of children on crowdfunding success would be significantly positive if the campaign was related to children. In addition, we confirm that the effect is indeed positive as it is consistent with the regression model. 
\end{itemize}

\begin{table}[H]
\centering 
  \caption{Counterfactual analysis experiment for categories.} 
  \label{ccategory} 
\begin{tabular}{@{\extracolsep{4pt}}lcccc} 
\\[-1.8ex]\hline 
\hline \\[-1.8ex] 
 & \multicolumn{4}{c}{\textit{Year}} \\ 
\cline{2-5} 
\\[-1.8ex] & \multicolumn{2}{c}{2020} & \multicolumn{2}{c}{2019} \\ 
\\[-1.8ex] & Predict & Ratio &Predict & Ratio\\ 
\hline \\[-1.8ex] 
Travel \&   Adventure       & 3389 & .0141 & 2159 & .0296 \\
Environment                & 3342 & .0000        &      &     \\
Babies Kids \& Family       & 3322 & -.0060 & 2077 & -.0095 \\
Sports Teams \& Clubs       & 3346 & .0012 & 2091 & .0029 \\
Competitions \& Pageants    & 3369 & .0081 & 2172 & .0358 \\
Non-Profits \& Charities    & 3332 & -.0030 & 2097 & .0000  \\
Medical Illness \& Healing  & 3177 & -.0494 & 1994 & -.0491 \\
Volunteer \& Service        & 3371 & .0087 & 2093 & -.0019 \\
Business \& Entrepreneurs   & 3343 & .0030 & 2104 & .0033 \\
Weddings \& Honeymoons      & 3357 & .0045 & 2120 & .0110 \\
Funerals \& Memorials       & 3249 & -.0278 & 2058 & -.0186  \\
Missions Faith \& Church    & 3338 & -.0012  & 2080 & -.0081 \\
Education \& Learning       & 3292 & -.0150 & 2058 & -.0186  \\
Other                      & 3323 & -.0057 & 2095 & -.0010 \\
Animals \& Pets             & 3403 & .0183 & 2122 & .0119 \\
Celebrations \& Events      & 3432 & .0269  & 2137 & .0191 \\
Creative Arts Music \& Film & 3350 & .00240 & 2087 & -.0048 \\
Accidents \& Emergencies    & 3299 & -.0129 & 2071 & -.0124  \\
Rent Food \& Monthly Bills  & 3342 & .0000        &      &   \\
Dreams Hopes \& Wishes      & 3343 & .0003 & 2096 & -.0005 \\
Community \& Neighbors      & 3341 & -.0003  & 2093 & -.0019 \\
\hline 
\hline \\[-1.8ex] 
\end{tabular} 
\end{table}

\begin{table}[H] \centering 
  \caption{Counterfactual analysis experiments for important features.} 
  \label{fcategory} 
\begin{tabular}{@{\extracolsep{5pt}}lcccc} 
\\[-1.8ex]\hline 
\hline \\[-1.8ex] 
 & \multicolumn{4}{c}{\textit{Year}} \\ 
\cline{2-5} 
\\[-1.8ex] & \multicolumn{2}{c}{2020} & \multicolumn{2}{c}{2019} \\ 
\\[-1.8ex] & Predict & Ratio &Predict & Ratio\\ 
\hline \\[-1.8ex] 
Goal          & 7    & -.9979 & 91   & -.9577 \\
Shares        & 3315 & -.0157 & 2006 & -.0665 \\
Donors        & 4784 & .4204 & 3603 & .6766 \\
days          & 3392 & .0071 & 2180 & .0144 \\
have kid      & 3392 & .0071 & 2228 & .0368 \\
Black         & 3024 & -.1021 & 1938 & -.0982 \\
Asian        & 3250 & -.0350 & 2100 & -.0228 \\
White        & 3421 & .0236 & 2162 & .0310 \\
happy         & 3447 & .0235 & 2199 & .0233 \\
anxiety           & 3348 & -.0059 & 2138 & -.0051 \\
text sad      & 3268 & -.0297 & 2139 & -.0047 \\
text scores   & 2047 & -.2853 & 1665 & -.2252 \\
technical scores   & 3451 & .0249 & 2177 & .0130 \\
\hline 
\hline \\[-1.8ex] 
\end{tabular} 
\end{table}

\section{LDA Topic Generation and Analysis}

\begin{figure*}
    \centering
	\includegraphics[width=1\textwidth]{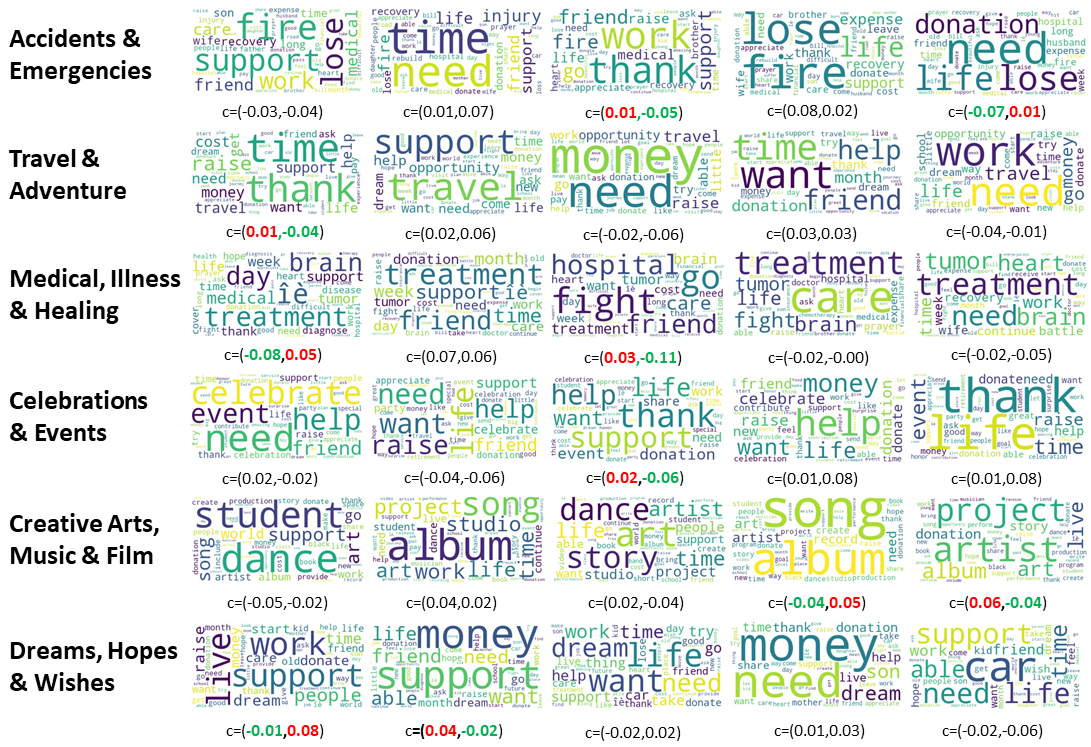}
	\caption{Topics under different categories. A red value indicates that the topic has a positive impact on success, while a green value indicates that the topic has a negative impact on success. Note those topics with changes in the polarity of the impact before and after the COVID-19 pandemic. }
	\label{topic}
\end{figure*}

We further analyze the relationship between crowdfunding success and the topics in each category by using LDA (Latent Derelicht Allocation) topic modeling\cite{blei2003latent}. For each category, we employ LDA to divide it into five topics. We build a regression model to analyze which topics have the most positive impact on the success rate of crowdfunding and which topics have the most negative impact. We focus on analyzing the categories that experience significant changes before and after the COVID-19 outbreak.
We focus on \emph{Accidents \& Emergencies}, \emph{Travel \& Adventure}, \emph{Medical, Illness \& Healing}, \emph{Celebrations \& Events}, \emph{Creative Arts, Music \& Film and Dreams}, and \emph{Hopes \& Wishes}. The results are shown in the Figure~\ref{topic}, where c is the correlation coefficient between success and topics. The first value represents the correlation coefficient before the COVID-19 outbreak, while the second value represents the correlation coefficient after the COVID-19 outbreak. We find that the frequencies of some words in the same category are very high, resulting in a high overlap between different topics. Therefore, we add the words whose frequency is in the top ten of all topics to the stop words list. Next, we apply an LDA model again to generate the topics and repeat this process until there is no such word. In the end, the coherence of the topic model is 0.34803. In addition, we find that there are some significant changes in the \emph{Other} category before and after the COVID-19 outbreak. However, the \emph{Other} category includes various topics, hence we separately construct a topic model for this category to analyze the impact of such topics on the success rate of crowdfunding.

\begin{figure*}\centering
	\includegraphics[width=1\textwidth]{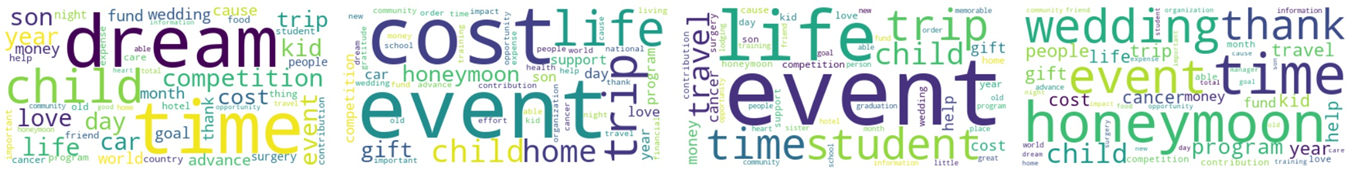}
	\centering
	\caption{Topics under the \emph{Other} category that had significant effect on success before the COVID-19 outbreak. We generate four topics with the best coherence value: 0.35192.}
	\label{other2019}
\end{figure*}

\begin{figure*}\centering
	\includegraphics[width=0.5\textwidth]{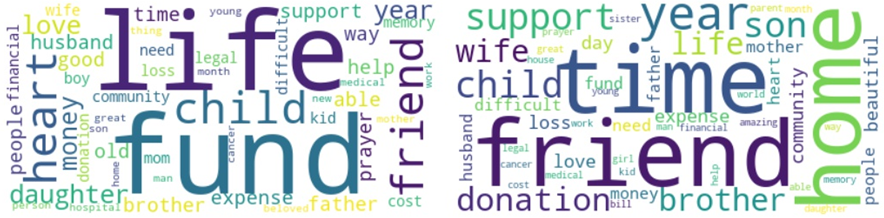}
	\centering
	\caption{Topics under the \emph{Other} category that had significant effect on success during the COVID-19 pandemic. We generate two topics with the best coherence value: 0.32625.}
	\label{other2020}
\end{figure*}

We find that there are some significant changes in the \emph{Other} category before and after the COVID-19 outbreak. We analyze the the \emph{Other} category individually because it contains various topics. We start with a topic number with the best coherence value. We find that before the COVID-19, topics in the \emph{Other} categories are relatively scattered, and we divide them into four topics(coherence score: 0.35192, as shown in Figure ~\ref{other2019}). Interestingly, topics are relatively concentrated after the COVID-19 outbreak as we can only divide the category into two topics (coherence score: 0.32625, as shown in Figure~\ref{other2020}). We find that before  COVID-19, the \emph{Other} category includes dreams, gifts, children, travel, Honeymoon \& Wedding, and other topics. Among these topics, combined with mentioned results, we find that most of these topics have a negative impact on the success of crowdfunding. However, after the COVID-19 outbreak, most topics focus on family, children, friends, and medical care. These topics have a positive impact on the success of crowdfunding. This suggests that the significance of the \emph{Other} category has changed because of the change of topics under the \emph{Other} category before and after the COVID-19 outbreak.

\section{Discussions and Conclusions}

This study has three limitations. The first limitation is the number of crawled campaigns. Second, we cannot control for every possible factor, and the nature of this study design leaves the possibility of residual confounding. Finally, we did not consider some dynamic factors, such as the change in the amount of a single donation. 

In conclusion, our study  analyzes the changes of significant features impacting crowdfunding success before and after the COVID-19 outbreak and validates three hypotheses. The study results suggest a substantial difference in some categories between before and after the COVID-19 outbreak. While \emph{dreams, travel}, or \emph{other} topics are less likely to be funded before the COVID-19 outbreak,  people began to make donations to these campaigns after the outbreak of COVID-19. People began to pay more attention to \emph{medical, accident}, and \emph{charity}. Some categories have not changed before or after the COVID-19 outbreak. For example, campaigns including \emph{babies, family, funerals}, and \emph{memorials} have always been easier to get people's donations. In contrast, campaigns such as \emph{sports, weddings}, and \emph{missions} have not been easy to get donations. More importantly, we find that there are significant differences in crowdfunding success among different races. The COVID-19 has made it more challenging for Black and Asian to raise money because the COVID-19 pandemic exacerbated such existing social disparities.

Our study can also guide and support fundraisers and crowdfunding companies like GofundMe to raise money. For example, we find that both the emotion of the text and emotion of the cover image impact the success of crowdfunding. In addition, we find that sadness, anxiety, or anger have no effect or even negative effect on crowdfunding. Compared with these negative emotions, positive emotions are more likely to get funded. Therefore, when the fundraisers write the description or upload the cover image, in addition to describing their misfortune and difficulties, they should also express more about their optimistic attitudes towards life. Furthermore, our success prediction model can help the company provide feedback scores to fundraisers, making it easier for fundraisers to succeed. In the future, we will examine a longer period of campaigns, as opposed to the 2-year data employed in this study. In addition, we will take other dynamic factors, such as each donation into consideration.

\bibliographystyle{IEEEtran}
\bibliography{mybibfile}

\end{document}